\documentclass[
    ,final            
  ]
  {aipproc}

\layoutstyle{6x9}

\begin{document}

\title{Measuring the NuMI Beam Flux for MINERvA}


\author{Melissa T. Jerkins}{
  address={ 
Department of Physics, The University of Texas at Austin\\ 
\vspace{4 mm}
\emph{On behalf of the MINERvA Collaboration}\footnote{http://minerva.fnal.gov}
  }
}



\begin{abstract}
MINERvA is employing multiple tools to understand its neutrino beam flux. We utilize external hadron production data, but we also depend heavily on \emph{in situ} techniques in which we reduce our hadron production uncertainties by tuning our Monte Carlo to both MINERvA detector data and muon monitor data. 
\end{abstract}

\maketitle


\section{Introduction}

\begin{figure}
  \includegraphics[height=.2\textheight]{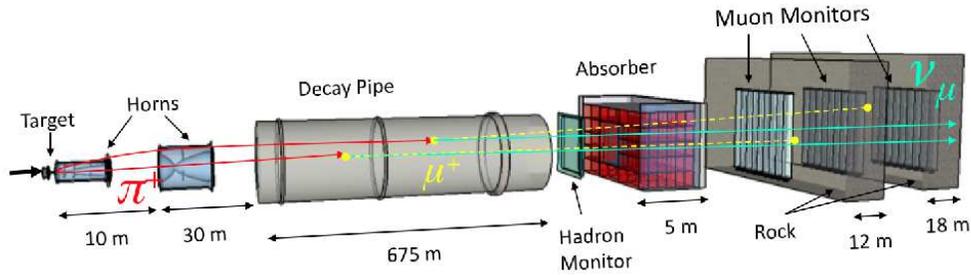}
  \caption{Diagram of Fermi National Laboratory's NuMI beamline preceding the MINERvA detector~\cite{Zarko2008}}
  \label{fig:beamline}
\end{figure}

MINERvA is performing low energy precision cross-section measurements using the Neutrinos at the Main Injector (NuMI) beamline at Fermi National Laboratory. Figure~\ref{fig:beamline} shows the NuMI beamline, which begins with 120~GeV protons hitting a carbon target and producing secondary pions and kaons. Magnetic focusing horns focus the hadrons, which proceed into a long decay pipe where they decay to muons and neutrinos. A hadron monitor measures the hadron content of the beam prior to an absorber through which only muons and neutrinos exit. Three muon monitors sit within excavated alcoves, and muons range out in the rock so that the beam contains almost exclusively neutrinos when it encounters the MINERvA detector. In order for MINERvA to make absolute cross section measurements, we need to understand our neutrino beam flux, which is notoriously difficult to measure. Because of the importance of the flux determination, MINERvA utilizes three different tools in order to understand the flux. We use existing hadronic cross section data to predict our yield of pions and kaons off of our target, but we also emphasize the importance of \emph{in situ} flux measurements, particularly beam fitting over both MINERvA detector data and muon monitor data.

\section{Sources of Flux Uncertainties}
\begin{figure}
  \includegraphics[height=.25\textheight]{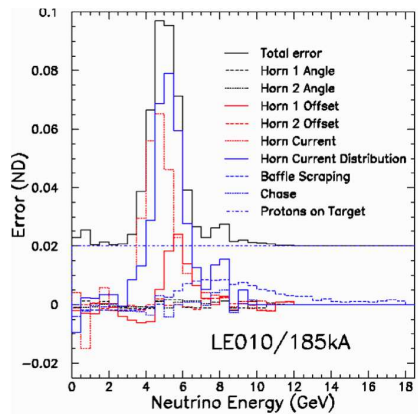}
  \caption{Relative error introduced into the flux determination by beam focusing uncertainties~\cite{Zarko2008}}
  \label{fig:focusing_errors}
\end{figure}

The most significant flux uncertainties come from hadron production uncertainties, but some are due to beam focusing. NuMI utilizes horn focusing to increase its flux, and Figure~\ref{fig:focusing_errors} shows a simulation of the uncertainties that result from inevitable misalignments in the focusing elements~\cite{Zarko2008}. The uncertainties are quite small in the focusing peak, and they are largest in the falling edge of the peak. These misalignments are fairly easy to understand and model using Monte Carlo techniques, and the effects of correlations between the various sources of uncertainty are being studied.

The dominant flux uncertainties derive from uncertainties in hadron production off of the target, and existing Monte Carlo models are not in good agreement with each other concerning how to model hadron production. For example, Monte Carlo models do not agree on the average transverse momentum of pions leaving the NuMI target, and those discrepancies result in different neutrino spectra because the pions are focused differently by the horns. 

\section{External Hadron Production Data}
Given that modern hadron production data sets continue to improve, MINERvA is certainly utilizing this information to learn about its flux. MINERvA must acknowledge, however, that we cannot depend entirely on experiments like NA49~\cite{na492007} to simply deliver our flux measurement. Most hadron production data sets are taken on thin targets, but the NuMI target is approximately two interaction lengths, meaning that reinteractions are a non-negligible 20-30\% effect. Even if we had a data set that corresponded perfectly to our target, we would still have to account for in-beam temporal variations that change the flux. We also have to account for downstream interactions where hadrons are created by interactions in material other than the target.

\section{Beam Fits to MINERvA Detector Data}
\begin{figure}
  \includegraphics[height=.2\textheight]{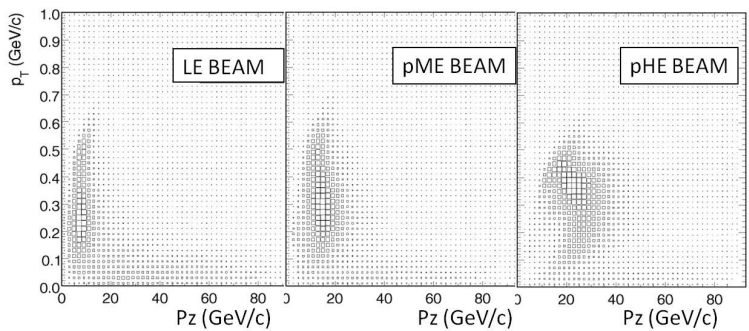}
  \caption{Distribution in $p_{z}$ and $p_{T}$ of hadron parents that yield a neutrino that interacts in the detector for the low energy, pseudo-medium energy, and pseudo-high energy beam configurations~\cite{Kopp2007}}
  \label{fig:pz_pt}
\end{figure}

To address these effects MINERvA wants to make an \emph{in situ} flux measurement that will average over real effects in the beam, and specifically we want to determine the flux as a function of the underlying hadron parent $p_{z}$ and $p_{T}$. The hadron momentum is kinematically related to the resulting neutrino energy, and the transverse momentum of the hadron determines how well the hadron is captured by the focusing horns. Figure~\ref{fig:pz_pt} shows a distribution of the hadron parents that yield a neutrino that interacts in the detector, and it illustrates that different NuMI beam configurations focus particles from different regions in the parent hadron $p_{z}$-$p_{T}$ space~\cite{Kopp2007}.

MINERvA intends to utilize the flexible design of the NuMI beamline. In addition to varying the current in the focusing horns, we can also move our target in and out of the first horn, which changes which hadrons are focused. This technique is not as effective as actually changing the separation between the focusing horns, but it is much faster and easier to accomplish. By producing neutrinos of the same energy using several different beam configurations, we are able to deconvolve focusing effects, hadron production off of the target, and neutrino cross sections. Because of these flexible beam configurations, we can attempt to tune our hadron production yields to match data from the MINERvA detector~\cite{Kopp2007,Kostin2001}. Each $p_{z}$-$p_{T}$ bin contributes with a different weight in each beam configuration. We create a functional form to parameterize our Monte Carlo yields off the target in terms of $p_{z}$ and $p_{T}$, and we then warp the parameters so that the Monte Carlo matches the real data. We perform a chi-squared minimization across multiple different beam configurations to learn how to tune our Monte Carlo.

The results of such a fit are a set of weights that should then be applied to Monte Carlo pion and kaon yields. MINOS successfully lowered their flux uncertainties using this technique, although they used an inclusive charge current data sample, which is not suitable for deconvolving neutrino cross sections.


\subsection{``Standard Candle'' Sample}
MINERvA wants to perform its fit on a ``standard candle'' data sample in which the cross section is approximately constant with neutrino energy. Ultimately we will perform our fit over several different standard candle samples, one of which will be quasi-elastic events of moderate $Q^{2}$. The energy independence of the cross section for this sample does not depend on axial mass. Low $Q^{2}$ events are excluded because of uncertainties due to nuclear effects, and high $Q^{2}$ events are excluded because of reconstruction difficulties as well as because of a non-neglible M$_A$ dependence. MINERvA will determine the shape of its flux by performing a global beam fit on our quasi-elastic sample. We will fix our overall normalization by using our inclusive charge current sample above approximately 20~GeV and comparing that to data from experiments like CCFR, CCFRR, and CDHSW~\cite{ccfr,ccfrr,cdhsw}, which measured the cross section in that region to within $\sim$3\%. 

\subsection{Estimate of Flux Error Bands}
\begin{figure}
  \includegraphics[height=.25\textheight]{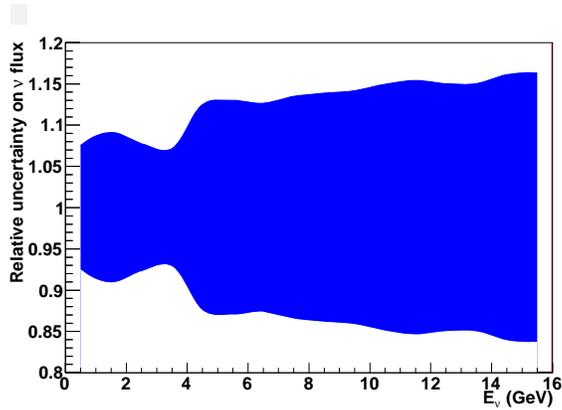}
  \caption{Error band applied to a low-energy neutrino flux that represents our uncertainty before tuning our Monte Carlo. Uncertainties due to kaon production and tertiary production are not included.}
  \label{fig:enuhist_stacked_norm}
\end{figure}

While MINERvA continues to collect ``special run'' data using NuMI's flexible design to map out the parent $p_{z}$-$p_{T}$ space, we are examining the size of the flux uncertainties that we expect to obtain after we perform a global beam fit. Before we attempt any hadron production fitting, we want to know what the state of ``current knowledge'' is. The uncertainties in current knowledge can be assessed by comparing hadron production models of various Monte Carlo simulations and observing the different neutrino fluxes they yield. Figure~\ref{fig:enuhist_stacked_norm} shows an error band applied to a low-energy neutrino flux that represents our uncertainty before tuning our Monte Carlo. It includes model differences in pion production, beam focusing uncertainties, and an overall 5\% yield uncertainty for $\pi^{+}$ production off of the target. It does not include model differences for kaons off of the target, which would primarily affect the high energy tail of the spectrum, and it does not include uncertainties due to downstream interactions, which are being studied.

We would like to know how much those uncertainties could be reduced by performing a hadron production tuning fit, but until MINERvA has taken and analyzed all of its special run data, we cannot actually perform such a fit. A study is underway, however, to estimate what the results of that fit might yield by warping the hadron production model in Geant4~\cite{geant4} Monte Carlo until it matches the neutrino fluxes yielded by Fluka~\cite{fluka} Monte Carlo. The assumption is that our underlying parameterization of hadron production off the target is flexible enough that the uncertainties in the fit will be similar when we actually perform it on real data.

\section{Beam Fits to Muon Monitor Data}
In addition to fitting MINERvA detector data to better understand our flux, MINERvA also utilizes \emph{in situ} muon monitoring, which is our second key tool for measuring our flux. The NuMI beamline preceding the MINERvA detector contains four excavated rock alcoves, three of which are instrumented with arrays of ionization chambers, and we plan to instrument the fourth alcove soon to increase our sensitivity. Each of our three arrays of ionization chambers are filled with helium gas. The muons ionize the gas, and we detect the resulting electrons. Significant amounts of rock separate the monitors, meaning that each monitor is sensitive to a different muon energy threshold, which translates into a different neutrino energy threshold. This type of muon monitoring is a proven technique that has been used for decades in both flux measurements and beam diagnostics, and NuMI's flexible beam design only increases the utility of these measurements~\cite{Kopp2006}.

We can tune our Monte Carlo hadron production model to match muon monitor data in the same way we tune it to match MINERvA detector data. We acquire data in the muon monitors from various beam configurations, and we perform a global fit to warp our underlying hadron production model in our Monte Carlo so that it agrees with the data. 

The data points over which we perform the fit are integrals of muon monitor fluxes from different beam configurations, and we correct the data for changes in variables like ambient temperature and pressure. In order to tune our Monte Carlo using this data, we also have to apply some corrections to our Monte Carlo data. Most importantly, we have to apply an overall scale factor to account for the charge we expect in the monitor per muon. We obtain this factor from beam tests done on prototype chambers. Unfortunately this factor is very sensitive to gas impurities and can vary 5-10\% even due to a contamination as small as 20~ppm. Once we have made the necessary corrections to both data and Monte Carlo, then we can tune our Monte Carlo to match the muon monitor data taken across multiple beam configurations. MINOS performed such a fit in which they floated the $\pi^{+}$ parameters and fixed the $\pi{+}/\pi^{-}$ and $\pi^{+}/K^{+}$ ratios using existing hadron production data, and they successfully obtained a flux shape measurement using muon monitor data~\cite{Loiacono2010}. 
Many backgrounds had to be estimated in that analysis, which contributed a significant amount of uncertainty to the fit and unfortunatley forced MINOS to normalize their flux measurement using near detector data.
MINERvA intends to repeat this analysis, but we hope to reduce some of the dominant uncertainties, particularly those related to gas impurities and backgrounds from $\delta$-rays. We are attempting to improve the purity monitoring of the gas, as well as our Monte Carlo modeling of $\delta$-ray production. 


\subsection{Modeling $\delta$-Ray Production}
$\delta$-rays are knock-on electrons created by muons as they travel through rock, air, or even the monitors themselves. These electrons can penetrate the monitor, ionize the helium gas, and result in a background signal. We expect the energy deposited by $\delta$-rays to increase with muon momentum and decrease with the amount of air in front of the monitors. Summing over all of the muons that reach each alcove, the Monte Carlo predicts that $\delta$-rays can contribute as much as 30\% of the muon monitor signal, meaning that accurate Monte Carlo modeling of this background is essential. 

To better understand $\delta$-ray production, we placed aluminum absorber plates in front of the monitors to deliberately introduce more $\delta$-rays into the data in a controlled way. By measuring the change in $\delta$-ray production with and without these various curtains of absorbers, we hope to constrain Monte Carlo $\delta$-ray production. A full analysis of this data is underway and will greatly improve the utility of the muon monitor fit.

\section{Conclusions}
MINERvA is employing multiple tools to measure, check, and cross-check its flux. We are gleaning what information we can from hadron production data, and we are depending heavily on \emph{in situ} techniques where we reduce our hadron production uncertainties by tuning our Monte Carlo to both MINERvA detector data and muon monitor data. By diversifying our measurement tools, we hope to lower our flux uncertainties from $\sim$30\% to 5-10\%.







\bibliographystyle{aipproc}   






\end{document}